\documentclass[12pt,preprint]{aastex}

\received{2013 September 12}
\accepted{2013 December 30}

\slugcomment{}
\shorttitle{Ups and Downs of $\alpha$~Cen}
\shortauthors{Ayres}

\begin{document}

\title{The Ups and Downs of Alpha Centauri}

\author{Thomas R.\ Ayres}

\affil{Center for Astrophysics and Space Astronomy,\\
389~UCB, University of Colorado,
Boulder, CO 80309;\\ Thomas.Ayres@Colorado.edu}

\begin{abstract}

The following is a progress report on the long-term coronal ($T\sim 1$~MK) activity of $\alpha$~Centauri A (HD\,128620: G2~V) and B (HD\,128621: K1~V).  Since 2005, {\em Chandra}\/ X-ray Observatory has carried out semiannual pointings on AB, mainly with the High Resolution Camera (HRC-I), but also on two occasions with the Low-Energy Transmission Grating Spectrometer (LETGS), fully resolving the close pair in all cases.  During 2008--2013, {\em Chandra}\/ captured the rise, peak, and initial decline of B's coronal luminosity.  Together with previous high states documented by {\em ROSAT}\/ and {\em XMM-Newton,}\/ the long-term X-ray record suggests a period of 8.2${\pm}$0.2~yr, compared to 11~yr for the Sun; with a minimum-to-peak contrast of 4.5, about half the typical solar cycle amplitude.  Meanwhile, the A component has been mired in a Maunder-Minimum-like low state since 2005, initially recognized by {\em XMM-Newton.}\/ But now, A finally appears to be climbing out of the extended lull.  If interpreted simply as an over-long cycle, the period would be 19.1${\pm}$0.7~yr, with a minimum-to-peak contrast of 3.4.  The short X-ray cycle of B, and possibly long cycle of A, are not unusual compared with the diverse (albeit much lower amplitude) chromospheric variations recorded, for example, by the HK Project.  Further, the deep low state of A also is not unusual, but instead is similar to the $L_{\rm X}/L_{\rm bol}$ of the Sun during recent minima of the sunspot cycle.

{\bf\small (Note: This preprint includes one additional {\em Chandra}\/ HRC-I pointing, in 2013 December, which was carried out after the final revision was submitted to the Journal.  The new X-ray points are consistent with the long-term trends and do not affect any of the quantitative conclusions.)}

\end{abstract}


\section{INTRODUCTION}

Alpha Centauri is a remarkable hierarchical triple star system, a little more than a parsec from the Sun.  The central binary consists of a solar-mass yellow dwarf (G2~V: ``A'') and a slightly less massive orange dwarf (K1~V: ``B''), separated by about 20~au; orbited at great distance ($\sim10^4$~au) by a dim, low-mass red dwarf (M6~V: ``C'' aka ``Proxima,'' nearest  star to the Sun).  Given the closeness of $\alpha$~Cen, an age only slightly older than the Sun, similar \ion{Ca}{2} HK activity, and bracketing the Sun in mass, AB are important solar analogs.  Characteristics of the system, and a history of its myriad observations, can be found in the comprehensive review by Beech (2012).  

The present report examines long-term changes in $\alpha$~Cen's coronal X-rays.  This is a follow-on to a previous study of coronal activity of AB (Ayres 2009; where additional details specifically relevant to the present work can be found).  Interest in the activity cycle of especially B has heightened recently with the proposed discovery of a close-in hot Earth-mass planet in orbit around the star (Dumusque et al.\ 2012; but see also Hatzes 2013).  The Doppler-reflex measurements were obtained when B was at the peak of its current cycle, and the increased spottedness of the K dwarf was a key systematic effect the authors had to confront in isolating the subtle radial velocity signal.

Robrade, Schmitt, \& Favata (2012) have described the X-ray history of the $\alpha$~Cen components through mid-2010, as part of a larger study of coronal cycles of solar-like dwarfs, based on measurements exclusively from {\em XMM-Newton.}\/  Unfortunately, the small -- and closing -- separation of AB (4.4$^{\prime\prime}$ at end of 2013) in the current arc of their 80~year orbit is challenging for $10^{\prime\prime}$-resolution {\em XMM-Newton,}\/ and its European Photon Imaging Camera (EPIC) has poor sensitivity for very soft coronal sources like $\alpha$~Cen A (see Ayres 2009).  The authors proposed a period of 8--9 years for the more cleanly observed (X-ray brighter) B component, and 12--15 years for fainter, blended A.  DeWarf, Datin, \& Guinan (2010) have investigated the long-term high-energy activity of specifically $\alpha$~Cen~B, incorporating additional tracers such as ``subcoronal'' ultraviolet fluxes.  They proposed a cycle of 8.8${\pm}$0.4~yr.  These estimates compare with Ayres' (2009) tentative nine year period for B, based on the aggregated X-ray measurements then available (2008 December), from {\em ROSAT, XMM-Newton,}\/ and {\em Chandra.}\/  

The present work extends the {\em Chandra}\/ time line by five years, showing a definitive minimum, rise, peak, and turn down of the $\alpha$~Cen B X-rays.  Together with previous high states seen sequentially by {\em ROSAT}\/ and {\em XMM-Newton,}\/ the new peak implies a well-defined cycle of $8.2{\pm}0.2$~yr for B over the past two decades.  Further, the sun-like A component finally appears to be emerging from an extended coronal low state (originally noted by Robrade, Schmitt, \& Favata [2005], and somewhat ominously described by them as the ``darkening of the solar twin'').

\section{OBSERVATIONS}

\subsection{High Resolution Camera}

{\em Chandra}\/ pointings on $\alpha$~Cen utilized the High-Resolution Camera (HRC), either in direct imaging mode (HRC-I); or recording the zeroth-order spatial image, and 5--175~\AA\ soft X-ray spectrum, with the Low-Energy Transmission Grating Spectrometer (LETGS) and HRC-S camera (``LETG0'' for the zeroth-order image)\footnote{LETG0/HRC-S is 6.7${\pm}$0.7 times less sensitive than HRC-I for soft coronal sources, based on paired HRC-I and LETGS pointings on $\alpha$~Cen in 2007 and 2011.}.  Thanks to {\em Chandra's}\/ 1$^{\prime\prime}$ resolution, there is little or no cross-talk between the AB event clouds (unlike {\em XMM-Newton}\/ for which the AB sources have been badly blended since 2005).  Characteristics of the instrument and general circumstances of the observations have been summarized by Ayres (2009).  Table~1 catalogs the full complement of {\em Chandra}\/ HRC-I and LETGS exposures to date.  

Figure~1 is a schematic streak image of the HRC pointings, from the ``first-light'' LETGS observation early in the {\em Chandra}\/ mission (Raassen et al.\ 2003) to the most recent HRC-I exposure in 2013 December.  The large proper motion of AB to the West dominates, but parallactic wobbles also can be seen, emphasized by the six-month cadence of the program and the large parallax ($0.75^{\prime\prime}$).  The slower orbital dance of the pair also is evident: AB are drawing toward a close approach on the sky in 2016, with a minimum apparent separation of only $4^{\prime\prime}$.  HRC captures the hierarchical, and in some instances subtle, motions of the system with a typical residual displacement of only 0.5$^{\prime\prime}$ (relative to the Pourbaix et al.\ [2002] ephemeris and Luyten [1976] proper motion); testament to the high accuracy of the {\em Chandra}\/ aspect reconstruction.  

There are several improvements in the current work over the earlier Ayres (2009) study.  First, an average X-ray flux was extracted from each HRC event list by filtering out periods of apparent flare activity.  This is especially relevant to the more recent pointings on B, during the peak of its starspot cycle when flares are more common.  Second, the X-ray light curves were corrected for dead time effects.  These were ignored in the previous study, and generally are not important (the dead time fraction normally is $<1$\%).  However, closer examination uncovered one instance where the dead time fraction was unusually high ($\sim$50\%) for the entire observation (ObsID~6375: causing a spurious $L_{\rm X}$ dip in the earlier Ayres [2009] time series); a second which had a cluster of dead time spikes in the middle of the sequence (ObsID~10980); and a few other, albeit more minor, anomalies. Third, as described in more detail later, it has been possible to use the LETGS spectra of $\alpha$~Cen, and two other low- to moderate-activity coronal stars, to undertake a  recalibration of the HRC-I energy conversion factors (ECF), which translate count rates into energy fluxes in a specific bandpass.

Figure~2 (modeled after Fig.~4 of Robrade et al.\ 2012) concatenates all the {\em Chandra}\/ pointings, with 300~s binning for the HRC-I events and 2100~s for LETG0.  Counts were collected in a $r= 1.6^{\prime\prime}$ detect cell, centered on the source, corresponding to 95\% encircled energy (EE).  An average background was determined in a $30^{\prime\prime}{\le}r{\le}90^{\prime\prime}$ annulus, as measured from the geometrical center of the binary in each epoch.  In all cases the background, scaled to the tiny detect cell, was less than 1\% of the (fainter) A signal.  The pipeline dead time corrections (2.1~s cadence) were averaged in the same way as the source counts, and divided into the binned light curves.  

These X-ray measurements have high precision, because many events are accumulated into the averages for each epoch; but potentially low accuracy, because short-term coronal fluctuations -- due to flares, rotational modulation of inhomogeneous surface structures, or active region emergence and decay -- can add significant systematic bias when a long-term trend is measured only infrequently with snapshot observations.  This fundamental sampling issue is addressed quantitatively later.

Within each time series, a mean level was determined from the main body of the count rate (CR) distribution, ignoring any high-CR tail resulting from transitory flare activity.  The light curves in Fig.~2 indicate that B was experiencing heightened flare activity as it rose in X-ray luminosity between 2010--2013, as noted by Robrade et al.\ (2012); whereas A only recently has begun to climb out of an extended X-ray minimum, still displaying few, if any, discernible coronal transients.  The obvious B flares typically are relatively brief compared with the $\sim$10~ks exposures, allowing a clear reading of the ``non-flare'' level.

The resulting flare-filtered CRs of AB are listed in Table~1, together with X-ray luminosities taking into account the distance and activity-dependent ECFs described later.  

\subsection{Low Energy Transmission Grating Spectrometer}

A new LETGS spectrum of AB was taken in mid-2011, when the secondary star was near the peak of its long-term coronal cycle.  The 2011 exposure, and previous two LETGS epochs (late-1999 and mid-2007), are illustrated schematically in Figure~3.  The plus and minus arms of the spectrum were folded onto each other, and divided by the effective exposure time at each merged wavelength, taking into account the detector gaps.  Also displayed are two comparison stars from the {\em Chandra}\/ archive: low-activity Procyon ($\alpha$~CMi; HD\,61421: F5~IV-V; ObsID's 10994 and 12042) and moderate-activity $\epsilon$~Eridani (HD\,22049: K2~V; ObsID 1869).  (These comparison spectra were used in the recalibration of the HRC-I ECF described later.)  The 2007 AB spectrum was especially instructive.  As noted by Ayres et al.\ (2008), $\alpha$~Cen A lacked the normal high-energy features in the ``iron L-shell'' region (1--2 keV), but still had significant emissions longward of 30~\AA.  This offered an explanation why {\em XMM-Newton}\/ saw such a dramatic ``fainting'' of the solar twin:  the EPIC cameras must use a thick blocking filter to suppress visible photon contamination from bright optical sources like $\alpha$~Cen, and the filter potentially could cut out more of the soft X-rays than anticipated. ({\em Chandra's}\/ HRC is a different design specifically immune to ``red leak'' and does not require a blocking filter or other extraordinary measures for bright stars.)  

The 2011 spectrum finds A still in a low state (i.e., mainly lacking the Fe L-shell features), while B had brightened up significantly below 20~\AA\ compared with 2007, now rivaling $\epsilon$~Eri (in appearance, although lower in $L_{\rm X}/L_{\rm bol}$).  

The LETGS spectra have value in two important ways.  First, calibrated line fluxes can be modeled using the differential emission measure (DEM) approach (as described by Ayres [2009] for the two earlier $\alpha$~Cen LETGS exposures, and those of the comparison stars mentioned above; and Raassen et al.\ [2003] for the first-light [late-1999] LETGS spectrum of $\alpha$~Cen).  A DEM model provides insight into the distribution of material with temperature in the stellar corona, which in turn is related to the global heating and cooling processes that shape the hot outer atmosphere.  Second, the detailed X-ray spectral distribution of an object plays a key role in establishing an appropriate ECF (erg~cm$^{-2}$~count$^{-1}$) to translate an HRC-I count rate into an energy flux.  The existence of an LETGS spectrum, in effect, compensates for the lack of innate energy discrimination in the HRC detectors.  To determine an ECF, an X-ray spectral energy distribution (SED) of the object is required: either the observed LETGS spectrum itself, or an SED calculated from a best-fit DEM model.  The first option is preferred, if there is sufficient signal-to-noise in the observed spectrum, because even the most comprehensive line emissivity tabulations (e.g., the Astrophysical Plasma Emission Database [APED]: Smith et al. 2001) still are missing many of the numerous weaker features, and occasionally the underlying atomic physics data are found to be inaccurate for even the stronger species (e.g., Beiersdorfer et al.\ 2002).  (The most recent version of APED [2.0.2: Foster et al.\ 2012] was utilized in the DEM modeling that follows.)

A related issue is that of instrument cross-calibration.  The ECF calculation (e.g., eq.~5 of Ayres 2009) is a ratio of integrals in which the SED appears in both numerator and denominator.  Thus, the ECF is first-order independent of the absolute level of the SED, caring mainly about the relative spectral distribution.  The absolute part of the ECF comes from the energy-dependent effective area curve of the instrument.  Nevertheless, if the LETGS and HRC-I are properly calibrated, the integral of the resolved LETGS flux densities over some reference energy band, say 0.2--2~keV, should equal the energy flux derived from a point-source HRC-I measurement fluxed with the corresponding ECF.  This is an important issue here, because of the starkly different results obtained for $\alpha$~Cen A, in its low state, with HRC-I compared with the {\em XMM--Newton}\/ EPIC imagers.  It is possible that the higher fluxes obtained by {\em Chandra}\/ might result from an inaccurate low-energy calibration of HRC-I (see Robrade et al.\ 2012), as opposed to the Ayres et al.\ (2008) suggestion that the {\em XMM--Newton}\/ thick-filter transmission instead might be at fault.

This issue was addressed by undertaking a recalibration of the LETGS spectra, as processed through the custom software (in IDL) utilized here.  The approach is described in general terms by Ayres (2009) and has the heritage of Beuermann et al.\ (2006), who promoted the value of hot white dwarfs, like HZ\,43, and the neutron star RXJ\,1856.6-3754 (hereafter RXJ\,1856), for calibrating the soft response of LETGS; and of Pollock (2004), who similarly advocated the value of power-law AGN sources for establishing the higher energy behavior of X-ray grating instruments.

To carry out the updated calibration, all the available LETGS pointings on HZ\,43, RXJ\,1856, and the blazars Mrk\,421 and PKS\,2155-304, from Y2000 to the present, were collected from the {\em Chandra}\/ archive, processed through the custom pipeline, and coadded (multiple observations were available for all the objects).  This yielded high-resolution distributions of count rates with wavelength for each object type.  The measured count rate densities ($C_{\lambda}$, in count s$^{-1}$ \AA$^{-1}$) are related to the (ideally  well-known, independently determined) true photon flux densities ($p_{\lambda}$, in pht cm$^{-2}$ s$^{-1}$ \AA$^{-1}$) by,
\begin{equation}
C_{\lambda}= {\cal R}_{\lambda\,\lambda^{\prime}} \otimes [g^{(1)}_{\lambda^{\prime}}\, (A_{\rm eff})_{\lambda^{\prime}}\, p_{\lambda^{\prime}}]\,\, ,
\end{equation}
where ${\cal R}$ is a redistribution matrix that describes the scattering of photons into higher diffraction orders relative to the first order; $g^{(1)}$ is the grating efficiency for first order; and $A_{\rm eff}$ (cm$^{2}$) is the effective area curve for all the instrumental components other than the grating assembly itself.  ${\cal R}$ differs from the more familiar ``response matrix,'' which for a grating instrument would have a nearly pure diagonal structure, with weaker adjacent off-diagonals reflecting the broadening of an input $\delta$-function source according to the wavelength dependent spectral point response function of the instrument.  Here, the higher-order redistribution matrix also is diagonally dominant (all diagonal elements are unity, since the redistribution coefficients are defined relative to first order), but contains far-off-diagonal elements fanning out from the origin ($\lambda= 0$) and located according to integer divisors of the input wavelength, $\lambda^{\prime}/m$ for order $m$, and populated by the relative grating efficiencies, $g^{(m)}_{(\lambda^{\prime}/m)}/g^{(1)}_{(\lambda^{\prime}/m)}$, reflecting the photons scattered into wavelength $\lambda$ from shorter wavelengths in the higher orders.  

The ${\cal R}$ matrix was constructed using revised values of the grating efficiencies for orders 1--25 in the file ``letgD1996-11-01greffpr001N0007.fits'' available from the {\em Chandra}\/ Calibration Database\footnote{see http://asc.harvard.edu/cal/letg/HO2011/}.  The fundamental wavelength scales (1--200~\AA) were built with a constant step of 25~m\AA, about three points per LETGS spectral resolution element.  A constant wavelength bin is crucial for the success of the matrix approach, because an unresolved emission line that subtends $\Delta\lambda$ \AA\ at wavelength $\lambda$ in first order, also will occupy the same $\Delta\lambda$ in any higher order $m$, because the resolution of the higher order increases as $m$, but so does the output wavelength, $m\,\lambda$, thus $\Delta\lambda$ remains constant.  In fact, the innate LETGS resolution in first order also increases in proportion to $\lambda$, thus instrumental-resolution emission features at their native first-order wavelengths are indistinguishable from any higher-order, also unresolved, features diffracted to those wavelengths.

Finally, given an independent measurement of the target X-ray absolute photon densities in the LETGS range, the effective area curve can be deduced from the observed $C_{\lambda}$ by dividing by the product ${\cal R}_{\lambda\,\lambda^{\prime}} \otimes [g^{(1)}_{\lambda^{\prime}}\, p_{\lambda^{\prime}}]$.  One can provide the $p_{\lambda^{\prime}}$ spectrum at the nominal wavelength-dependent LETGS resolution, to avoid having to introduce additional line-spread terms directly into ${\cal{R}}$; but this is irrelevant for the calibrators used here, which all are smooth continuum sources.  

Beuermann et al.\ (2006) tabulated consensus SEDs for HZ\,43 and RXJ\,1856.  The authors also provided a model for the latter based on a dominant $T\sim 32$~eV black-body plus a $T\sim 63$~eV ``hot spot,'' absorbed by an interstellar column, $N_{\rm H}\sim 1.1\times 10^{20}$ cm$^{-2}$.  More recently, Kaastra et al.\ (2009) published model atmospheres for HZ\,43; their ``model~2'' agrees best with the Beuermann et al.\ photon densities.  Finally, Pollock (2004) listed photon power law indices (describing the photon flux density in pht cm$^{-2}$ s$^{-1}$ keV$^{-1}$) and galactic columns for several blazars including Mrk\,421 and PKS\,2155-304.  The LETGS blazar spectra utilized here were coadded over various X-ray luminosity states, but the assumption was made that the average spectrum still could be modeled by a single photon power law.  Having the two sets of blazar spectra provided an independent check.  

The strategy was to let HZ\,43 define $A_{\rm eff}$ for $\lambda> 65$~\AA; use RXJ\,1856 to extend $A_{\rm eff}$ down below 40~\AA; then scale the blazars, which cover $\lambda<50$~\AA\ well, to match RXJ\,1856 in the interval of overlap.  In essence, RXJ\,1856 served as a transfer standard between the best understood spectrum -- that of the hot white dwarf -- and the well-characterized blazar power laws.  In fact, RXJ\,1856 matched the soft end of the HZ\,43 $A_{\rm eff}$ very well, although a small adjustment of $\sim$1\% was applied to the RXJ\,1856 curve based on the wavelengths in common with HZ\,43.  

In short, $A_{\rm eff}$ was stitched together over several different types of calibrators, but fundamentally tied to the white dwarf model fluxes.  Then, any arbitrary $C_{\lambda}$ spectrum can be translated to absolute photon flux densities according to the solution of eq.~1 for $p_{\lambda}$ (which includes a matrix multiplication of ${\cal R}^{-1}$ against the $C_{\lambda}$ spectrum to, in effect, delete the higher order photons).  The flux density, $f_{\lambda}$ (erg cm$^{-2}$ s$^{-1}$ \AA$^{-1}$), can be obtained trivially from $p_{\lambda}$ by multiplying by the energy per photon, $h\,c/\lambda$.

As part of the calibration process, the systematic $\sim 10$\% decline in HRC-S sensitivity\footnote{see: http://asc.harvard.edu/cal/Hrc/hrcsqeu\_201108.html; Note: HRC-I has not experienced a similar sensitivity decrease.} over the {\em Chandra}\/ mission was taken into account, as follows.  There are 14 LETGS spectra of HZ\,43, taken between early 2002 and early 2011, which were used in the LETGS calibration.  Once the full effective area curve was derived from the stitched-together calibrators, each of the individual HZ\,43 spectra then was fluxed, and integrated in several 15--30~\AA\ bands covering the wavelength region 55--155~\AA\ where the S/N of the white dwarf is high.  The bandpass fluxes displayed a systematic decline of $\sim 0.7$\% yr$^{-1}$, independent of wavelength, relative to the mean epoch of the calibration, 2004.8 (where the LETGS fluxes identically match the WD calibration model).  A linear correction in time then was incorporated in the LETGS fluxing procedure to remove the (albeit small) systematic error that otherwise would be present in the $\alpha$~Cen LETGS series (covering an 11~year span).

Figure~4 illustrates results of the radiometric calibration procedure for selected bright lines of $\alpha$~Cen AB from the three epochs of LETGS spectra.  Because the longwavelength ends of the two more recent AB observations are partially blended (e.g., at \ion{Fe}{9} $\lambda$171 and \ion{Fe}{10} $\lambda$174), the normal ``bow-tie'' extraction template was replaced with a one-sided bow, extending in the direction opposite to the other spectral stripe, and the extracted counts were adjusted for the reduction in collecting area.  Similarly, the off-spectrum background was collected in a one-sided manner, again opposite to the other spectrum in the cross-dispersion direction.  The bow-tie extraction template, itself, was determined from a super-coaddition of all the reference spectral images (white dwarf, neutron star, and the blazars), each registered in the cross-dispersion direction by centroiding.  The boundaries of the high-S/N super-coadd were traced such that at any wavelength, about 95\% of the counts were collected.  This is a compromise between maximizing the 1-D cross-dispersion ``encircled energy,'' while minimizing the background (which depends linearly on the width of the extraction template, and becomes increasingly a factor at the longer wavelengths where the bow broadens significantly).

Table~2 lists fluxes of representative LETGS lines that were incorporated in the DEM modeling described below.  The LETGS spectra divide, conveniently, into low and high states for both stars: epochs 2007\,+\,2011 for A-low, 1999 for A-high, 1999\,+\,2007 for B-low, and 2011 for B-high.  The pairs of LETGS spectra for the two sets of low states were averaged to improve S/N for those fainter epochs.  The specific features in the table were selected on the basis of line strength, relative freedom from apparent blends in the observed spectra as well as in the emissivities, good coverage of coronal temperatures, and representing both low- and high-FIP species\footnote{The solar corona, and stellar counterparts, tend to display systematically enhanced abundances of low first ionization potential species like Mg, Si, and Fe: see Feldman (1992)}.  The formation temperature listed represent a weighting by both the emissivities and the DEM model(s), so can differ significantly from the peak emissivity temperature (noted in Fig.~4) if, for example, the DEM is sharply peaked at a lower or higher temperature.  The features, themselves, were measured in the fluxed LETGS spectra using a Gaussian fitting procedure, or -- in cases where a feature was weak or absent -- numerical integration over the effective width of an instrumental-resolution feature.  Three-sigma upper limits were assigned if the measured flux did not achieve that significance.  

Despite the updated LETGS calibration effort, the new X-ray line fluxes generally are in good agreement with the comparable measurements reported in Ayres (2009), although somewhat higher (up to $\sim$20\%) in several cases.  Table~2 also provides far-ultraviolet (FUV) fluxes, of Li-like \ion{C}{4} $\lambda$1548, \ion{N}{5} $\lambda$1238, and \ion{O}{6} $\lambda$1031, for the low and high states of AB.  The FUV resonance lines form between 1--3$\times 10^{5}$~K.  They serve as a lower boundary condition for the DEM modeling, tying into the He-like and H-like CNO features of the LETGS region.  The FUV fluxes were measured from archival {\em Hubble}\/ Space Telescope Imaging Spectrograph (STIS) E140M-1425 echellegrams of AB, particularly those taken since 2010, which cover the high state of B and low state of A; the A high state was captured by STIS in 1999.  The B low state was inferred from earlier long term {\em IUE}\/ measurements (see, e.g., Ayres et al.\ 1995).  The \ion{O}{6} fluxes were based on archival {\em FUSE}\/ data, in some cases (i.e., activity states not represented by the available spectra) scaling from the other, longer wavelength, FUV features.  The exact values of the FUV fluxes are not crucial, however, since they serve mainly as a distant boundary condition for the DEM modeling, as described next.

\subsection{Differential Emission Measure Modeling}

The final step, as a prelude to evaluating the time history of the $\alpha$~Cen X-ray fluxes, was to derive ECFs for the range of HRC-I count rates recorded from the two stars.  A common way to derive the conversion factors is to utilize a synthetic spectrum, simulated either from an isothermal model or -- better yet -- from an emission measure distribution constructed to match, say, LETGS line fluxes.  A DEM model has value in its own right, for the insight it can provide on the structure of the stellar corona, so the second approach was followed here.  The modeling approach has been described in detail by Ayres (2009).  The present version has been updated for the new selection of reference X-ray features, and the availability of an improved set of line emissivities (APED\,2.0.2).  

Figure~5 summarizes the results of the DEM modeling exercise.  Observed and simulated fluxes were normalized both by the bolometric flux of each star, $f_{\rm bol}$, to remove the bias of the different stellar sizes, and by the $\log{T}$-integrated radiative power curve, $P_{\rm tot}$, of the particular feature to allow the disparate species, which cover a wide range of intrinsic emissivities, to be compared on a more-or-less common scale (see Ayres 2009).  Initially, a full spectrum at LETGS resolution was created at each APED temperature step, by summing up all the line emissivities, treated as Gaussian profiles of the appropriate resolution centered at the respective line wavelengths.  Then, the temperature-resolved spectra were individually integrated over the same wavelength band as the observed spectra for each specific target feature (e.g., $\pm$1\,FWHM for a Gaussian fit, or the wavelength bandpass for a numerically integrated measurement).  These temperature-resolved, species-dependent power curves were the fundamental atomic-physics input to the DEM modeling.

Calculated fits to the observed fluxes in Fig.~5 are separated according to high-FIP and low-FIP.  In all cases a low-FIP abundance enhancement of 2 was applied (see Ayres [2009]).  The high-FIP species for each star and activity state are closely reproduced by the illustrated DEM distributions, and the low-FIP species fall into place as well, given the uniform abundance enhancement.  The new models are qualitatively similar to those derived in the previous study.  They all show a characteristic deep minimum at temperatures just below $\log{T}= 6.0$~K; with coronal peaks at around $\log{T}\sim$6.1--6.3~K; but display progressively more material at hotter temperatures moving up the ladder from A-low to B-high; rather than, say, simply an increase in the peak-$T$ DEM with increasing activity.  

The A low state perhaps could be interpreted as a ``basal'' corona lacking any magnetic active regions (i.e., starspots and surrounding ``plage''), as is characteristic of the Sun during one of its decadal spot minima.  The A high state (and both B states) might then represent the situation of adding active regions to the basal circumstance: on the Sun, sunspot groups have notably higher coronal temperatures, 2--3~MK, than the $\sim$1~MK quiet corona.  The emission of the FUV ``Transition Zone'' species also rises with increasing activity (A-low to B-high), but more slowly than the coronal ($T>1$~MK) counterparts.  This is a well-known activity trend (e.g., Ayres et al.\ 1995), and undoubtedly also is tied to an increasing prevalence of active regions.
 
\subsection{Energy Conversion Factors}

Finally, the pieces are available to solve the ECF puzzle.  An HRC-I ECF can be calculated for each activity-specific state of AB, which applies to one or more of the individual LETGS spectra (e.g., there are two A-low and two B-low cases), by dividing the integrated spectral flux, say 6.2--62~\AA\ (corresponding to the familiar 0.2--2~keV ``{\em ROSAT}'' band), by a convolution of the HRC-I effective area with the photon spectrum, integrated over all wavelengths for which $A_{\rm eff}$ is non negligible\footnote{Here, the effective area file ``HRC-I\_ea\_2010-OCTver.dat'' was appropriated from the {\em Chandra}\/ CalDB}.  (The formalism applies specifically to a sensor, like HRC, lacking any energy discrimination.) 

The trick is to find an appropriate reference SED.  This is a challenge because a DEM-motivated APED model is not a perfect representation of the intrinsic stellar X-ray spectrum, owing to missing lines and/or incomplete atomic physics; but, an LETGS spectrum also is not a perfect representation of the intrinsic stellar SED, because unavoidable photon noise can masquerade as real signal, or noise can mask true spectral structure, such as a weak  but ubiquitous bremsstrahlung continuum.  In fact, it could be important to consider wavelengths beyond the nominal limits of an LETGS spectrum (5--175~\AA, for the custom extractions here), depending on how far to longer, or shorter, wavelengths the HRC-I effective area extends.  The tabulation provided by the CalDB starts at 1.1~\AA, and ends at 200~\AA, where the area is about 0.5\% of the peak and dropping toward longer wavelengths, but not precipitously.  An APED spectrum for a 1~MK soft source, like $\alpha$~Cen A, has significant coronal emission beyond 200~\AA, and interstellar absorption is not much of a help in attenuating the longer wavelengths for these very nearby stars.  Nevertheless, in the absence of additional information, the HRC-I effective area was assumed to vanish beyond 200~\AA.  Experiments showed that extrapolating $A_{\rm eff}$ out to 250~\AA, based on the apparent slope below 200~\AA, caused about a 3\% decrease in the AB ECFs; a small uncertainty, to be sure, in the overall scheme.

Recall that the HRC-I ECF depends mostly on the shape of the coronal SED, rather than its absolute level.  Thus, it is essential to empirically validate the cross-calibration of HRC-I relative to LETGS.  This requires finding pairs of LETGS and HRC-I pointings on soft coronal targets taken close enough in time that variability issues are minimized.  Unsurprisingly, a search of the {\em Chandra}\/ archive revealed very few suitable pairs.  It would be unusual to take an HRC-I exposure of a target bright enough to record an LETGS spectrum, unless it were for calibration purposes: LETGS not only provides diagnostically valuable energy resolution, but also spatial information through the zeroth-order image; rendering a companion HRC-I pointing, well, pointless.  Thankfully, however, the two recent LETGS spectra of $\alpha$~Cen were paired with HRC-I exposures relatively close in time (ObsIDs 7432\,+\,7433, $\Delta{t}\sim 16$~d; 12332\,+\,12333, $\Delta{t}\sim 1$~d); potentially suitable material for the cross-calibration piece of the ECF. 

Furthermore, there is a single HRC-I exposure of the presumably very constant neutron star RXJ\,1856 (actually several pointings, but only one -- ObsID 4288 -- was taken on-axis with the target in focus), to match against the collection of LETGS spectra used in the recalibration exercise.  RXJ\,1856 is especially important for the cross-calibration validation, because its black-body spectrum covers the key range 20--70~\AA, closely corresponding to the soft end of the 0.2--2~keV band, and matching the peak of the $\alpha$~Cen A low state spectrum.  Because the neutron star thermal emission falls almost exclusively within that relatively confined wavelength interval, there is no issue with out-of-band flux.  The HRC-I ECF calculated for RXJ\,1856 -- based on the theoretical model, which was normalized to the observed average LETGS flux distribution, and matches it very well -- is 6.5 (in units of $10^{-12}$ erg cm$^{-2}$ count$^{-1}$), compared with the ``neutral'' value 7.3 (in same units) for a constant source spectrum between 6.2--62~\AA\ and zero flux elsewhere.  Because the HRC-I $A_{\rm eff}$ is itself relatively flat between 5--70~\AA, the ECF responds mainly to the mean energy of the source photons.  For a spectrum tilted toward higher energies, like the blazers, the ECF will increase from the neutral value, whereas for a source sloping toward lower energies, like RXJ\,1856, the ECF will be lower.  These considerations also are affected by out-of-band flux, such as for the low-activity stars discussed here, because significant counts can be collected outside the reference band, particularly 65--200~\AA\ for soft APED-like sources, thus causing the bandpass flux to be `diluted,' implying a lower ECF (the ECFs for the $\alpha$~Cen stars are $\sim 4$, as shown later). 

Given the ECF calculated for RXJ\,1856, and the integrated flux in the reference band derived from the empirically normalized model, the predicted HRC-I count rate ($f_{0.2-2}\,/\,{\rm ECF}$) is 1.70~cps.  The observed CR is 1.82~cps, measured from ObsID~4288 (1.6$^{\prime\prime}$ detect cell, accounting for 95\% EE factor), implying that the theoretical ECF is too large by $\sim 7$\%.  That, in turn, would suggest that the HRC-I $A_{\rm eff}$ is understated (on average, in the 0.2--2~keV band) by the same amount.  This compares with the 6\% uncertainty in the HRC-I quantum efficiency at low energies cited in the {\em Chandra}\/ CalDB documentation\footnote{see: http://cxc.cfa.harvard.edu/cal/Hrc/QE/hrci\_qe\_N0008.html}.

Turning back to the more normal stellar coronal sources, there are legitimate concerns, as alluded earlier, over choosing for the reference SED a purely theoretical APED spectral model versus a purely empirical, observed LETGS spectrum.  To balance these concerns, a hybrid approach was taken.  First, a best-fit DEM was developed for each of the target examples (low and high states of AB, and including the coadded LETGS spectrum of soft source Procyon and the single LETGS exposure of more active $\epsilon$~Eri).  Then, the resulting APED spectral distribution was normalized to the observed LETGS tracing over the range 6.2--62~\AA\, considering only those wavelengths for which the observed flux densities exceeded 2\,$\sigma$ with respect to the assigned photometric errors.  Finally, the portions of the full LETGS spectrum that survived the 2\,$\sigma$-cut were spliced into the normalized model SED.  In this way, all the brightest lines of the empirical spectrum were preserved, even those that might not be well reproduced by the APED simulation; but the low flux, noisy regions of the empirical spectrum were replaced by the (noiseless) theoretical fluxes, including the faint but ubiquitous bremsstrahlung continuum and those weaker lines captured by APED (although certainly not a complete set of those features).  

Taking the APED simulation shortward of the lower effective limit of the LETGS, $\sim$5~\AA\ for these soft sources, and beyond the instrumental upper limit, $\sim$175~\AA, out to the 200~\AA\ cutoff of $A_{\rm eff}$, accounted for the possible influence of those ``missing'' parts of the soft X-ray spectrum.  ISM attenuation of the DEM-derived spectrum was included, with $N_{\rm H}= 0.4$ (in units of $10^{18}$ cm$^{-2}$) for $\alpha$~Cen AB; 0.8 for Procyon; and 0.6 for $\epsilon$~Eri (Redfield \& Linsky 2008).

The 2\,$\sigma$ cut is a balance between wanting to include as much of the empirical spectrum as possible, while minimizing wavelengths that might be compromised by spurious fluctuations.  Nevertheless, ECFs calculated with cuts at 1.5\,$\sigma$ or 2.5\,$\sigma$ differed from the 2\,$\sigma$ result by less than 3\% in all cases.  However, the hybrid spectrum ECFs were systematically $\sim$10\% lower than predicted by the purely DEM-derived APED SEDs for the $\alpha$~Cen stars, although nearly identical for Procyon and $\epsilon$~Eri.

Next, the hybrid ECFs were compared against an activity index, defined as the predicted HRC-I count rate (i.e., the denominator of the ECF ratio) divided by the bolometric flux of the star (2.87 for A, 0.96 for B, 1.82 for Procyon, and 0.105 for $\epsilon$~Eri [native units are $10^{-5}$ erg cm$^{-2}$ s$^{-1}$]).  The ECFs displayed a systematic trend with the reduced count rate, ${\rm cr}\equiv {\rm CR}/f_{\rm bol}$, characterized by a power law index of 0.16\,.

Then, ``empirical'' ECFs (i.e., dividing the bandpass integrated 2\,$\sigma$-cut hybrid flux by an HRC-I CR from a pointing close in time to the LETGS exposure) were constructed for the two pairs of $\alpha$~Cen spectra.  Compared to the derived power law, the empirical ECFs fell below by 6${\pm}$3\% (uncertainty is a standard error of the mean: three of the values were consistently $\sim 10$\% low, the other, a few percent high), similar to RXJ\,1856.  Recall that the theoretical/empirical validation of the ECFs relies on source constancy between the LETGS and HRC-I pointings; certainly a good assumption for RXJ\,1856, but less so for the $\alpha$~Cen stars.  Since the majority of the four AB offsets, and their average, agree with the more reliable 7\% deficit indicated by RXJ\,1856, the constant coefficient in the initial power law was reduced by 7\% yielding a consensus conversion factor,
\begin{equation}
{\rm ECF}\sim\,3.7\,{\rm cr}^{0.16}\,\, ,
\end{equation}
in the units mentioned earlier.  Compared with the earlier {\em Chandra}\/ study, these ECFs are systematically 20--30\% smaller for the various activity states of $\alpha$~Cen.  The empirical validation increases confidence that the LETGS and HRC-I fluxes now are consistent, and ultimately traceable back to the fundamental calibration of HZ\,43.  The new ECF subsequently was applied to the HRC-I measurements of AB. 

In the chain of argument that led to this result, essentially all of the uncertainties are of the systematic type, because random errors due to photon statistics are negligible owing to the high quality exposures of the original calibrators, as well as of $\alpha$~Cen AB and the other comparison coronal stars.  An example of a potential systematic error was the decision to adopt the small offset between theoretical and empirical ECFs indicated by the calibration target RXJ\,1856, and supported by three of the $\alpha$~Cen LETGS/HRC-I pairs (but not the fourth).  The potential size of this  systematic error could be as large as the 7\% adopted for the offset.  There are additional possible systematic errors, especially those associated with the LETGS calibration and the choice of the ``2\,$\sigma$-cut'' hybrid spectrum approach, which could contribute comparable levels of uncertainty.  If the systematics combine incoherently, the cumulative systematic error could be less than $\sim$15\%; but if the systematics reinforce each other, the total uncertainty could be larger.

\section{ANALYSIS}

Figure~6 summarizes the time history of the $\alpha$~Cen coronal emissions, including the previous {\em ROSAT}\/ era (in two cases -- 1996.13 and 1996.64 -- averaged over month-long campaigns with 1--2 day sampling)\footnote{New ECFs for {\em ROSAT}\/ HRI were derived using the hybrid spectra described earlier: 10.5 (in the standard units) for the A high state (applicable to all the HRI pointings on A); 11.0 for the B low state (for the lone observation in 1998); and 14.8 for the B high state (1995--97).  These ECFs are 10--20\% lower than derived in the earlier study.}; all the {\em Chandra}\/ measurements (HRC-I flare-filtered averages and LETGS integrated fluxes); and the {\em XMM-Newton}\/ $L_{\rm X}$'s published by the Hamburg group (Robrade et al.\ 2012).  A factor of 1.28 was applied to the latter to match the apparent {\em Chandra}\/ cycle of B, reflecting an -- albeit small -- lingering disagreement between the ECF's of the thick-filtered EPIC cameras compared to HRC.  To be sure, the discrepancy is smaller than proposed in the earlier Ayres (2009) study (which was more like a factor of 2), at least for $\alpha$~Cen B; and comparable to the potential systematic errors described above.  This certainly is a positive outcome for the updated LETGS/HRC-I cross-calibration.  Note, however, that even with the scaling to match B, the {\em XMM-Newton}\/ $L_{\rm X}$'s for A, between 2005--2010, still fall significantly below the {\em Chandra}\/ values, perhaps adding weight to the earlier suggestion that the {\em XMM-Newton}\/ thick-filter calibration might be an issue for very soft sources, like A's low state.  

The upper panel of the figure reports solar 0.2--2~keV X-ray luminosities over the past two decades obtained from the Flare Irradiance Spectral Model (FISM)\footnote{see: http://lasp.colorado.edu/lisird/fism} tabulations, daily values integrated over the 6.2--62~\AA\ reference bandpass and averaged in 81-day bins (3 solar rotations).  Error bars are standard deviations within each time bin and illustrate the average amplitude of variability associated with rotational modulations and active region evolution; solar flares are too modest, too short in duration, and too infrequent to significantly affect these $\sigma$'s.  A long-term average over the preceding three solar cycles is presented as well (see Ayres [2009] for details).

The depression of the {\em XMM-Newton}\/ A coronal luminosities relative to {\em Chandra}\/ perhaps can be likened to the dramatic contrast between {\em Yohkoh}\/ X-ray images of the Sun at cycle minimum (uniformly dark disk, only a few scattered bright points) compared with softer imagers like the {\em SoHO}/EIT \ion{Fe}{12} 195~\AA\ channel (`fuzzy green ball:' substantial, ubiquitous coronal emission still present at minimum).  Nevertheless, all the instruments, stellar and solar, should be able to report consistent fluxes for any specified reference band within their sensitivity range, if calibrated properly.  It is worth recalling that the {\em Chandra}\/ HRC microchannel-plate sensors were specifically designed to have excellent soft response, if only because at least HRC-S must record the LETGS spectrum with good sensitivity out to 175~\AA\ (well beyond the $\sim$35~\AA\ limit of the Reflection Grating Spectrometer on {\em XMM-Newton}\,).  At the same time, one must be somewhat wary of a non-energy-resolving detector, like HRC, that can collect significant counts from out-of-band photons (i.e., outside the 0.2--2~keV reference interval), because, for example, dominant flux from the apparently less volatile coronal emissions beyond $\sim 60$~\AA\ could mask a sudden drop in intensity at the shorter, already intrinsically fainter wavelengths.  That is where the LETGS spectra have provided an essential grounding point, to show explicitly how the $\alpha$~Cen SEDs evolve from the A low state to the B high state. 

Also illustrated in Fig.~6 are sinusoidal fits in $\log{L_{\rm X}}$ to the AB X-ray light curves, including for B -- but not A -- the scaled {\em XMM-Newton}\/ values.  As seen in the upper panel, the solar cycle shape is only roughly log-sinusoidal.  Despite that, the simple model provides a good match to the B X-ray time series, with apparent peaks in 1995--96 in the {\em ROSAT}\/ era, {\em XMM-Newton}\/ circa 2004, and now {\em Chandra}\/ in 2012.  The resulting period is $8.2{\pm}0.2$~yr, somewhat smaller than the earlier estimates.  (The period uncertainty was estimated by a Monte Carlo approach, exploiting the empirical ``snapshot variability'' described below.)  The X-ray minimum-to-peak contrast is about 4.5, roughly half the typical solar amplitude, although B is several times more active than the Sun in $L_{\rm X}/L_{\rm bol}$.  B's cycle is reminiscent of the K5~V star 61~Cygni A ([HD\,201091] P$\sim$7~yr; contrast$\sim$3: Robrade et al.\ 2012).

As for $\alpha$~Cen A, only a single peak possibly has been seen (1998--2000).  The long interval of low, nearly constant $L_{\rm X}$ might indicate a lack of cycling, as in a Maunder minimum; or a delayed rise, like that experienced by the Sun transitioning to contemporary Cycle~24.  Blindly applying the log-sinusoidal model suggests a period of $19.1{\pm}0.7$~yr, about 70\% longer than solar normal.  The minimum-to-peak contrast is smaller than B's, about 3.4.  The solar twin does appear to be recovering in $L_{\rm X}$ compared with 2005--2010, so it remains to be seen whether it reestablishes a sun-like cycle, or continues in a more leisurely coronal holding pattern.

The average deviation of the observed $L_{\rm X}$'s with respect to the best-fit log-sinusoidal model is an empirical estimate of the snapshot variability.  For the {\em ROSAT}\/ plus {\em Chandra}\/ epochs, the deviation of the A luminosities from the best-fit model was about 10\%, while for B, now including {\em XMM-Newton},\/ it was about 22\%.  Additionally, the standard deviation of the A luminosities over the two campaigns in 1996, of nearly daily visits by {\em ROSAT,}\/ was $\sim$13\%; while for B, $\sim$19\%.  The potential influence of these$\sim$10--20\% short-term coronal fluctuations on the $\sim400$\% long-term trends thus appears to be minor, supporting the semiannual sampling used here with {\em Chandra,}\/ and by the {\em XMM-Newton}\/ group in their broader survey.

\section{CONCLUSIONS}

The $\alpha$~Centauri stars are quite sun-like in their fundamental stellar properties, and at least B displays a solar-like X-ray activity cycle, although with a somewhat shorter period (a ``class 1'' variable in the notation of the \ion{Ca}{2} HK Project: Baliunas et al.\ 1998).  In contrast, $\alpha$~Cen A has been behaving as if in a Maunder minimum, or perhaps just has an over-long normal cycle (``class 3'' variable).  Even at the minimum of A's contemporary variation (or lack of same), its $L_{\rm X}$ (and $L_{\rm X}/L_{\rm bol}$) still is similar to the Sun's during the extended minimum at the conclusion of Cycle~23.  This supports the conclusion of Ayres et al.\ (2008) and Ayres (2009) that the ``fainting of the solar twin'' witnessed by {\em XMM-Newton}\/ perhaps is more related to calibration issues, than a true indication that A had fallen into an ultra-deep minimum, well below anything seen so far on the Sun.  If the latter were the case, it would suggest that the pure convectively driven surface dynamo of the star (on the Sun responsible for the ubiquitous quiet corona, outside of active regions, present even at sunspot minimum, and relatively constant over the cycle: e.g., S\'anchez Almeida \& Mart\'inez Gonz\'alez [2011], and references therein) had at least partly failed.  This certainly is a possibility, given our present incomplete knowledge concerning the inner workings of dynamo processes; but also not a theoretical path that necessarily needs to be pursued, given {\em Chandra's}\/ alternative, less catastrophic, view of the $\alpha$~Cen A fainting episode.

With regard to the putative Earth-mass companion of $\alpha$~Cen B, the next few years should find the K dwarf approaching a minimum of its X-ray cycle.  The decreasing spottedness would mitigate one of the key systematic errors that affects the subtle Doppler-reflex measurements.  Countering that advantage, AB will be passing through a mutual close approach on the sky, increasing scattered light contamination of B's visual spectrum by brighter A.

In short, both components of $\alpha$~Cen appear to be exhibiting symptoms of coronal variability seen historically on the Sun through the lens of sunspot counts (and on other stars through HK monitoring): regular cycles mixed with occasional extended minima featuring few if any spots.  As such, $\alpha$~Cen AB -- with the most detailed stellar X-ray histories to date -- will continue to serve as keystone examples of the coronal counterparts of starspot cycles, to balance against the more accessible, but generally more subtle, chromospheric HK records.

\acknowledgments

This work was supported by grants GO1-12014X and GO2-13018X from the Smithsonian Astrophysical Observatory.  Observations from {\em Chandra}\/ X-ray Observatory were collected and processed at the CXO Center, operated for NASA by SAO.  This research made use of public databases hosted by {SIMBAD}, maintained by {CDS}, Strasbourg, France.  The author also thanks the Leibniz-Institut f\"ur Astrophysik Potsdam for their hospitality during a visit in which part of this study was conducted.

\clearpage
\begin{deluxetable}{rcrccccc} 
\rotate
\tabletypesize{\small}
\tablenum{1} 
\tablecaption{{\em Chandra}\/ HRC Pointings} 
\tablecolumns{8}
\tablewidth{0pt} 
\tablehead{\colhead{ObsID} &  \colhead{UT Mid-Exposure} & \colhead{$t_{\rm exp}$} & 
\colhead{$({\rm CR})_{\rm A}$}  &  \colhead{$({\rm CR})_{\rm B}$}  &  
\colhead{$(L_{\rm X})_{\rm A}$}  &  \colhead{$(L_{\rm X})_{\rm B}$}  & \colhead{Notes}\\ 
\colhead{}   & \colhead{(yr)} & \colhead{(ks)} & \multicolumn{2}{c}{(counts s$^{-1}$)} & 
\multicolumn{2}{c}{($10^{27}$ erg~s$^{-1}$)} & \colhead{}\\
\colhead{(1)} & \colhead{(2)} & \colhead{(3)} & \colhead{(4)} & \colhead{(5)} & \colhead{(6)} & \colhead{(7)} & \colhead{(8)}} 
\startdata 
\cutinhead{HRC-I}
    6373  &  2005.81   &   5.15  &    0.46${\pm}$0.04  &  2.20${\pm}$0.15     &     0.27   &  2.02  &  \\ 
    6374  &  2006.36  &    5.11   &   0.44${\pm}$0.03  &  1.01${\pm}$0.08     &     0.26   &  0.82  &  \\   
    6375  &  2006.96  &    2.67   &   0.41${\pm}$0.04  &  0.96${\pm}$0.09     &     0.24  &   0.77  & large dead-time correction  \\   
    7433  &  2007.47  &    5.04  &    0.43${\pm}$0.05  &  0.71${\pm}$0.04     &     0.26  &   0.54  &  \\  
    7434  &  2007.96   &   5.11  &    0.47${\pm}$0.04  &  0.75${\pm}$0.04     &     0.28  &   0.58  &  \\   
    8906  &  2008.39  &   10.08  &    0.47${\pm}$0.03  &  0.81${\pm}$0.09     &     0.28  &   0.64  & small B flare \\  
    8907  &  2008.96  &    9.34  &    0.47${\pm}$0.05  &  0.86${\pm}$0.06     &     0.28  &   0.68  &  \\   
    9949  &  2009.41  &   10.06  &    0.44${\pm}$0.04  &  1.48${\pm}$0.06    &      0.26  &   1.28  &  \\   
    9950  &  2009.95  &   10.05  &    0.49${\pm}$0.04  &  1.82${\pm}$0.08     &     0.30  &   1.62  &  \\  
   10980  &  2010.34  &    9.76  &    0.62${\pm}$0.06  &  3.32${\pm}$0.66     &     0.39  &   3.26  & dead-time spikes; large B flare \\  
   10981 &   2010.81  &   10.03  &    0.49${\pm}$0.05  &  2.76${\pm}$0.20    &      0.30  &   2.63  & small B flare \\   
   12333  &  2011.44   &   4.88  &    0.65${\pm}$0.06  &  2.17${\pm}$0.16    &      0.41  &   1.99  & small B flare  \\   
   12334  &  2011.99  &   10.07  &    0.56${\pm}$0.04  &  3.36${\pm}$0.17    &      0.35  &   3.30  & small B flare  \\   
   14191  &  2012.47  &   10.10  &    0.76${\pm}$0.07  &  2.82${\pm}$0.10   &       0.50  &   2.69  &  \\  
   14192  &  2012.95   &  10.06  &    0.93${\pm}$0.06  &  2.39${\pm}$0.10   &       0.62  &   2.22  &  \\  
   14193  &  2013.48  &   10.59  &    0.83${\pm}$0.08  &  1.94${\pm}$0.10    &      0.55  &   1.75  &  \\
   14232  &  2013.96   &  10.05  &    0.93${\pm}$0.06  &  2.00${\pm}$0.10   &       0.62  &   1.81  &  \\[3pt]   
 \cutinhead{LETGS}
    0029  &  1999.98   &  79.57   &   $\{^{0.148{\pm}0.009}_{2.86}$ &  
                                      $\{^{0.142{\pm}0.008}_{3.37}$ &    0.62  &   0.73  &  \\  
    7432  &  2007.43  &  117.08   &   $\{^{0.075{\pm}0.007}_{1.32}$ &  
                                      $\{^{0.104{\pm}0.009}_{2.44}$ &   0.28  &   0.53  &  \\  
   12332  &  2011.44  &   78.45   &   $\{^{0.097{\pm}0.007}_{1.83}$ &  
                                      $\{^{0.283{\pm}0.017}_{8.89}$ &    0.40  &   1.93  &  \\[3pt]
\enddata 
\tablecomments{Exposure time (col.~3) as reported in {\em Chandra}\/ archive, including dead-time correction.  ``Non-flare'' count rates (cols.~4 and 5) refer to a $r=1.6$\arcsec\ detect
cell, and were corrected for the 95\% EE.  Uncertainties reflect standard deviations of binned count rates with respect to reported flare-filtered averages.  In LETGS count rate columns, upper entry is for the zeroth-order image (LETG0), lower entry is a calibrated flux (0.2--2~keV: 10$^{-12}$ erg cm$^{-2}$ s$^{-1}$) derived from the ``hybrid'' LETGS spectrum as described in text.  X-ray luminosities (0.2--2~keV; cols.~6 and 7) were derived from the count rates using source-dependent ECF's (except for the LETGS values for which the integrated fluxes were used directly).  $(L_{\rm X})_{\odot}\sim$~0.2--1.5 in same energy band and luminosity units.} 
\end{deluxetable} 

\clearpage
\begin{deluxetable}{lccccc} 
\tablenum{2} 
\tablecaption{LETGS Line Fluxes} 
\tablecolumns{6}
\tablewidth{0pt} 
\tablehead{\colhead{ID~~$\lambda$~(\AA)} &  \colhead{$\log{T}$~(K)} & 
\multicolumn{2}{c}{$\alpha$~Cen~A} & \multicolumn{2}{c}{$\alpha$~Cen~B}\\[5pt]
  &  &  \colhead{Low State} & \colhead{High State} & \colhead{Low State} & \colhead{High State}\\[5pt] 
\multicolumn{1}{c}{ObsIDs:} &  &\colhead{7432\,+\,12332} & \colhead{29}  & \colhead{29\,+\,7432}  & \colhead{12332}
    } 
\startdata 
\cutinhead{High-FIP Species}
\ion{Ne}{ 9}~13.45 &  6.3--6.5 & ${\lesssim}2.6$ & ${\lesssim}3.3$ & $2.0{\pm}0.4$ & $14{\pm}1$ \\ 
\ion{ O}{ 8}~16.01 &  6.4--6.5 & ${\lesssim}1.8$ & ${\lesssim}1.3$ & $1.1{\pm}0.3$ & $9.8{\pm}1.1$ \\ 
\ion{ O}{ 7}~18.63 &  6.2--6.3 & ${\lesssim}1.1$ & ${\lesssim}1.6$ & $1.8{\pm}0.4$ & $5.3{\pm}0.8$ \\ 
\ion{ O}{ 8}~18.97 &  6.3--6.4 & ${\lesssim}1.4$ & $4.8{\pm}0.5$ & $8.2{\pm}0.4$ & $46{\pm}1$ \\ 
\ion{ O}{ 7}~21.60 &  6.2--6.3 & $3.4{\pm}0.5$ & $10{\pm}1$ & $14{\pm}1$ & $41{\pm}1$ \\ 
\ion{ O}{ 7}~22.10 &  6.2--6.3 & $3.5{\pm}0.5$ & $8.0{\pm}0.6$ & $9.9{\pm}0.4$ & $30{\pm}1$ \\ 
\ion{ N}{ 7}~24.78 &  6.2--6.3 & ${\lesssim}1.1$ & $1.5{\pm}0.5$ & $2.8{\pm}0.4$ & $8.9{\pm}0.8$ \\ 
\ion{ C}{ 6}~28.47 &  6.2--6.3 & ${\lesssim}1.1$ & $1.4{\pm}0.5$ & $1.1{\pm}0.3$ & $2.9{\pm}0.7$ \\ 
\ion{ N}{ 6}~28.79 &  6.2 & $1.7{\pm}0.4$ & $3.0{\pm}0.5$ & $2.2{\pm}0.3$ & $5.5{\pm}0.8$ \\ 
\ion{ N}{ 6}~29.54 &  6.2 & $1.1{\pm}0.3$ & $2.4{\pm}0.6$ & $1.3{\pm}0.4$ & $3.6{\pm}0.7$ \\ 
\ion{ C}{ 6}~33.73 &  6.2--6.3 & $6.3{\pm}0.3$ & $8.8{\pm}0.5$ & $8.6{\pm}0.3$ & $23{\pm}1$ \\ 
\ion{ C}{ 5}~40.27 &  6.1--6.2 & $4.3{\pm}0.6$ & $4.4{\pm}0.9$ & $3.1{\pm}0.6$ & $4.5{\pm}1.2$ \\ 
\ion{ O}{ 6}~150.1 &  5.8--5.9 & $4.9{\pm}0.5$ & $3.9{\pm}0.6$ & $1.9{\pm}0.5$ & $2.4{\pm}0.7$ \\ 
\ion{ O}{ 6}~173.1 &  5.8--5.9 & $3.7{\pm}0.8$ & $5.3{\pm}1.3$ & $5.7{\pm}1.0$ & ${\lesssim}5.4$ \\ 
\ion{ O}{ 6}~1031.9 &  5.5--5.6 & $84{\pm}10$ & $95{\pm}10$ & $62{\pm}12$ & $93{\pm}10$ \\ 
\ion{ N}{ 5}~1238.8 &  5.3 & $28{\pm}1$ & $29{\pm}1$ & $16{\pm}1$ & $26{\pm}1$ \\ 
\ion{ C}{ 4}~1548.2 &  5.1 & $185{\pm}1$ & $209{\pm}1$ & $95{\pm}1$ & $143{\pm}1$ \\ 
\cutinhead{Low-FIP Species}
\ion{Fe}{17}~15.01 &  6.4--6.6 & ${\lesssim}2.1$ & $2.7{\pm}0.5$ & $3.1{\pm}0.4$ & $29{\pm}1$ \\ 
\ion{Fe}{17}~17.07 &  6.4--6.6 & $2.1{\pm}0.7$ & $2.5{\pm}0.6$ & $4.8{\pm}0.4$ & $37{\pm}1$ \\ 
\ion{Ca}{11}~30.45 &  6.2--6.3 & ${\lesssim}1.6$ & ${\lesssim}1.5$ & $1.7{\pm}0.3$ & $4.2{\pm}0.7$ \\ 
\ion{Si}{11}~52.29 &  6.2--6.3 & $1.1{\pm}0.3$ & $2.6{\pm}0.5$ & $2.8{\pm}0.3$ & $6.2{\pm}0.6$ \\ 
\ion{Mg}{10}~57.88 &  6.2 & $1.9{\pm}0.3$ & $3.6{\pm}0.5$ & $2.8{\pm}0.3$ & $7.2{\pm}0.6$ \\ 
\ion{Si}{ 8}~61.04 &  6.1 & $7.7{\pm}0.4$ & $11{\pm}1$ & $5.4{\pm}0.4$ & $6.0{\pm}0.9$ \\ 
\ion{Fe}{ 9}~103.6 &  6.1 & $2.9{\pm}0.3$ & $3.0{\pm}0.5$ & $2.1{\pm}0.3$ & $3.1{\pm}0.5$ \\ 
\ion{Fe}{ 9}~105.2 &  6.1 & $2.9{\pm}0.3$ & $2.5{\pm}0.4$ & $1.8{\pm}0.3$ & $1.7{\pm}0.4$ \\ 
\ion{Ni}{11}~148.4 &  6.1  & $11{\pm}1$ & $15{\pm}1$ & $11{\pm}1$ & $17{\pm}1$ \\ 
\ion{Fe}{ 9}~171.1 &  6.1 & $106{\pm}1$ & $119{\pm}2$ & $81{\pm}1$ & $99{\pm}2$ \\ 
\ion{Fe}{10}~174.5 &  6.1 & $94{\pm}4$ & $54{\pm}4$ & $68{\pm}3$ & $109{\pm}6$ \\[3pt] 
\enddata 
\tablecomments{
~Line fluxes are in $10^{-14}$ erg cm$^{-2}$ s$^{-1}$.  Identifications are dominant emissivity at that wavelength, from APED~2.0.2.  Formation temperatures are typical values weighted by product of emissivity and DEM. Upper limits are 3\,$\sigma$ with respect to assigned measurement error.} 
\end{deluxetable} 

\clearpage
\begin{figure}
\figurenum{1}
\vskip  0mm
\hskip  -15mm
\includegraphics[scale=0.75,angle=90]{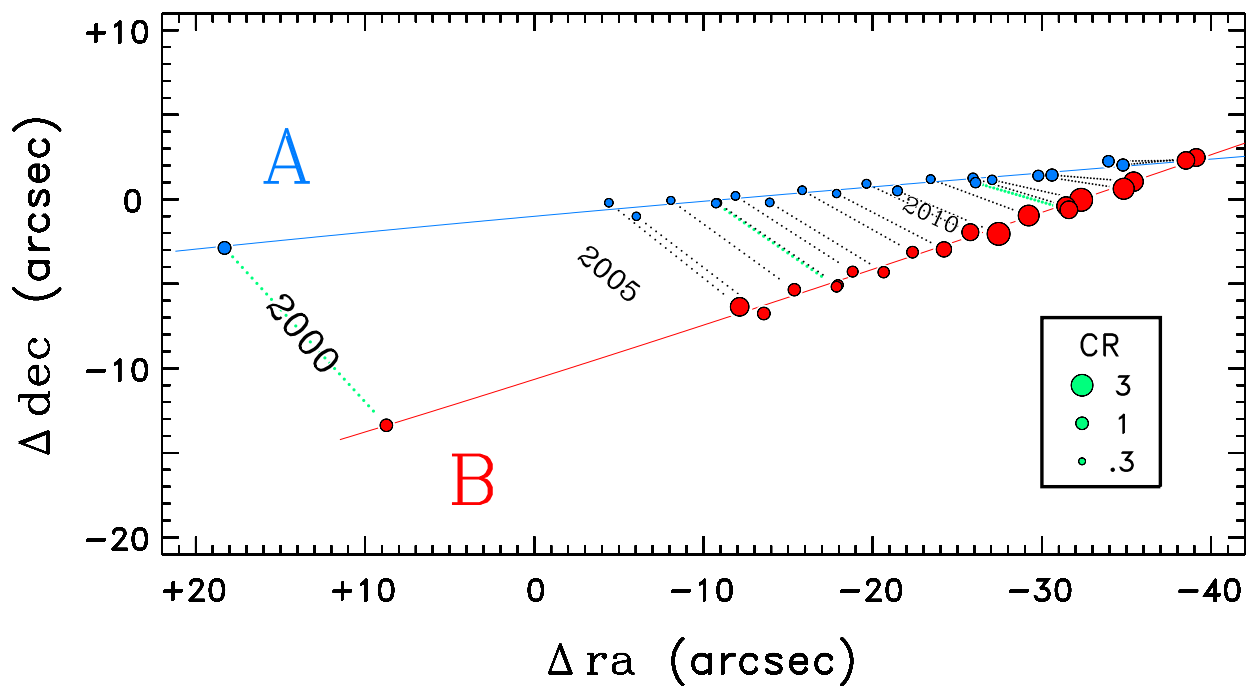} 
\vskip 0mm
\figcaption[]{{\em Chandra}\/ pointings on $\alpha$~Cen AB: North is up; West to the right.  Circles (blue for A; red for B) indicate HRC count rates, according to legend at right (LETG0 CRs multiplied by 7 to put them roughly on HRC-I scale).  Positions reflect centroids of the source event clouds, with no post-facto refinement of the astrometery.  Dotted lines connect AB in each epoch; heavier green dots indicate LETGS pointings.  Systematic drift to West is due to large proper motion of the binary; ``wobbles'' reflect 0.75$^{\prime\prime}$ annual parallax; while slower orbital dance ($P= 80$~yr) also is evident.  Thin solid curves are predicted AB trajectories:  O--C deviations are only 0.5$^{\prime\prime}$ on average.
}
\end{figure}

\clearpage
\begin{figure}
\figurenum{2}
\vskip  0mm
\hskip  -8mm
\includegraphics[scale=0.75,angle=90]{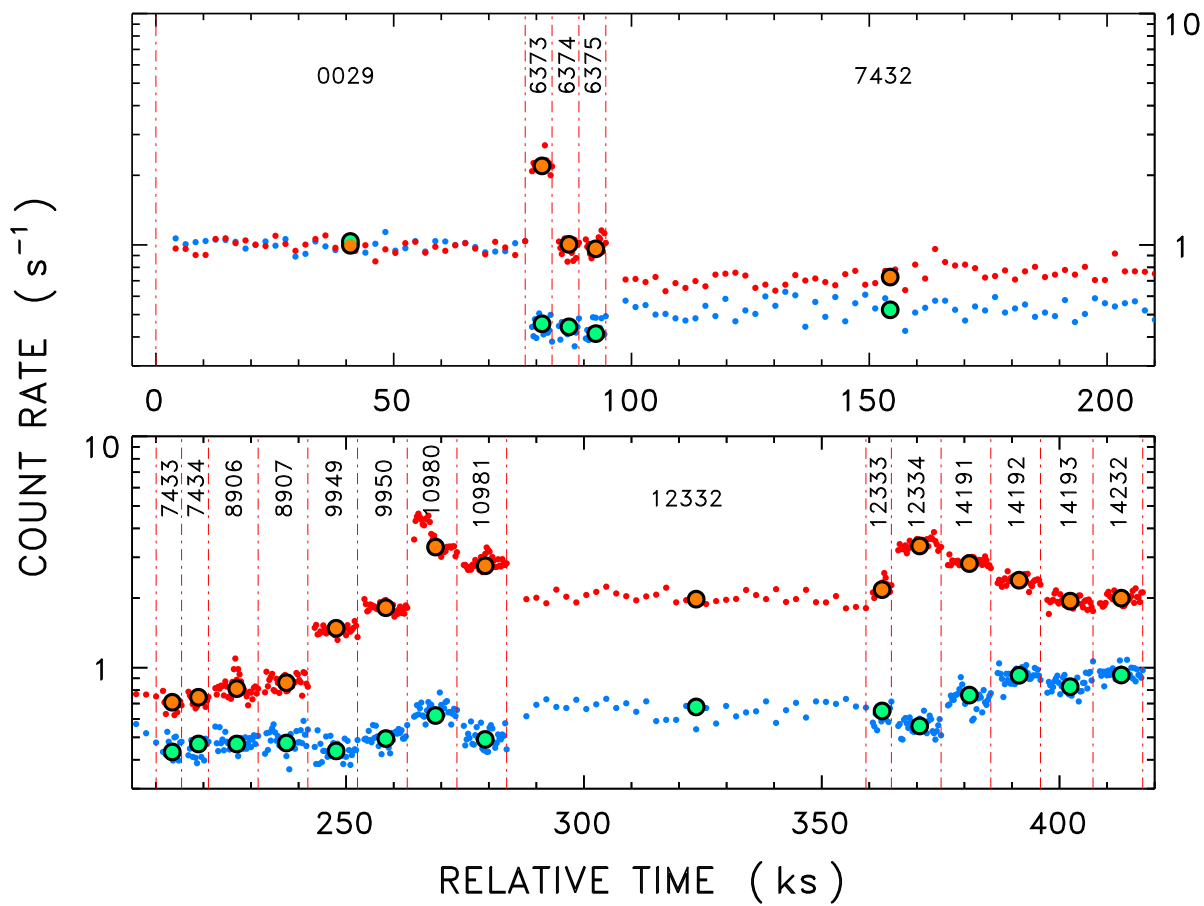} 
\vskip 0mm
\figcaption[]{{\em Chandra}\/ HRC light curves, labeled by ``ObsID'' (see Table~1).  Abscissa illustrates duration of exposures (LETGS are longest): blue points are for G-type primary; red, for K-type secondary; larger circles are ``flare-filtered'' averages.  LETG0 values again multiplied by 7.  Individual points also were corrected for temporally-dependent dead time effects and detect cell EE.
}
\end{figure}

\clearpage
\begin{figure}
\figurenum{3}
\vskip  0mm
\hskip  -6mm
\includegraphics[scale=0.625,angle=90]{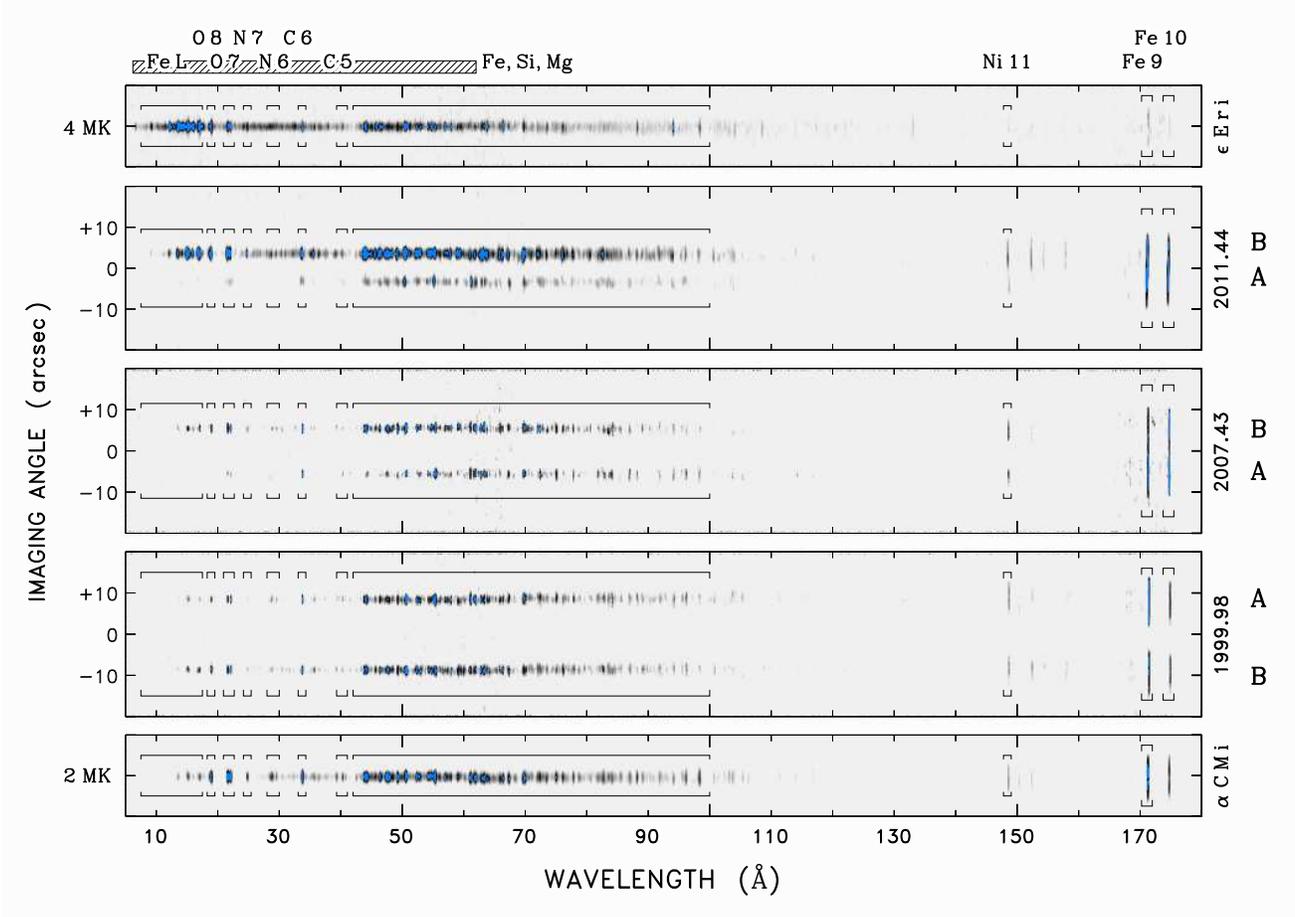} 
\vskip 0mm
\figcaption[]{LETGS spectral images (see Ayres et al.\ 2008).  Darker colors indicate higher CR's.  Middle three panels: $\alpha$~Cen AB in late-1999, when the two stars were nearly equal; mid-2007, when B was similar to 1999, but A had faded at higher energies (continuation of ``fainting'' episode witnessed by {\em XMM-Newton}\,); and mid-2011, when B was near the peak of its cycle, and A was beginning to recover from a long-term coronal lull.  Despite decreasing orbital separation in recent years, the AB spectral stripes are cleanly resolved, except at the longest wavelengths ($\sim$170~\AA) where they are partially blended in 2007 and 2011.  Alpha~CMi (bottom panel: $T\sim 2$~MK) and $\epsilon$~Eri (top panel: $T\sim 4$~MK) are low- and moderate-activity comparisons, respectively.  At upper edge of figure, hatched bar indicates 0.2--2~keV ``{\em ROSAT}'' energy band, and locations of key features are marked (``O\,8''= \ion{O}{8}, etc.). 
}
\end{figure}

\clearpage
\begin{figure}
\figurenum{4}
\vskip 0mm
\hskip -12mm
\includegraphics[scale=0.75,angle=90]{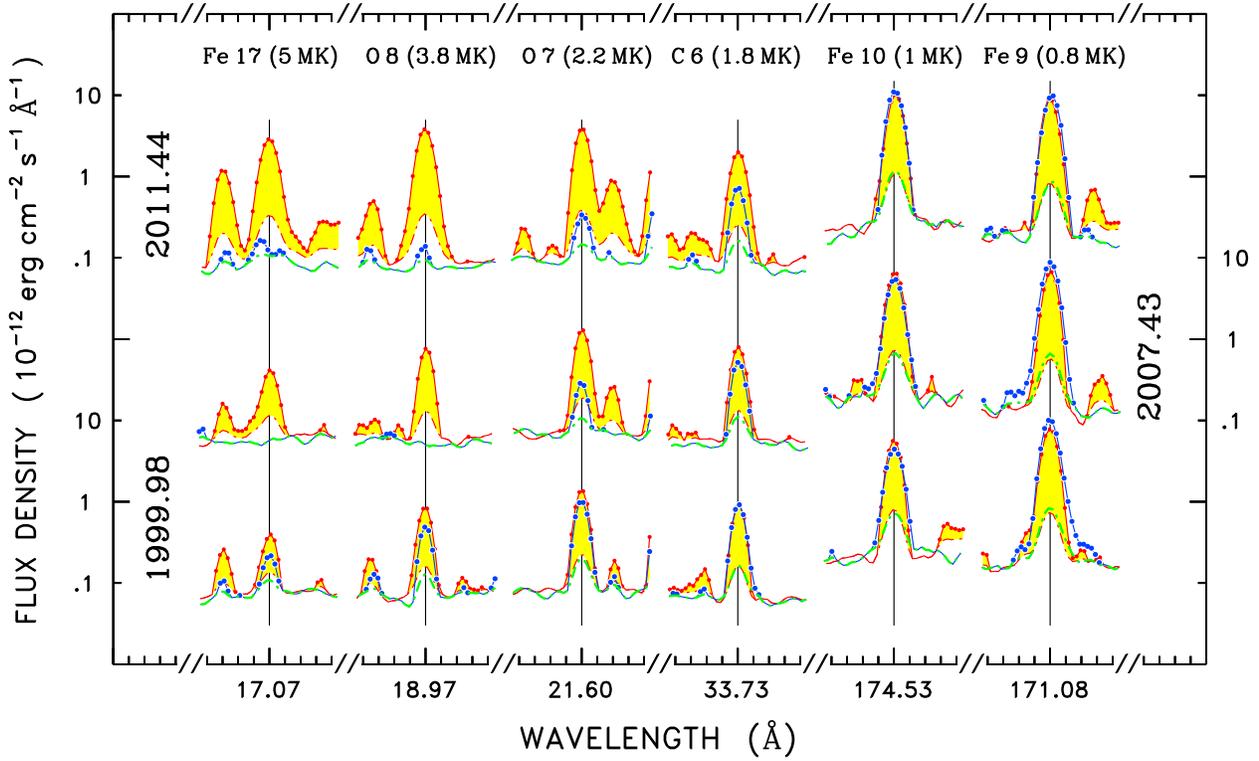} 
\vskip 0mm
\figcaption[]{Spectra of representative features from the three LETGS epochs.  K star profiles are yellow shaded outlined by red dots; G star is represented by blue dots.  Smoothed photometric errors (1\,$\sigma$) are orange and green dot-dashed curves, respectively.  Formation temperatures noted in upper portion of panel are ``peak emissivity'' temperatures: actual formation temperatures can differ significantly if the maximum of the emission measure distribution is at a much lower, or much higher, temperature.  Note dramatic differences between the two stars at shortest wavelengths ($\lambda< 20$~\AA), whereas differences are hardly noticeable at longest wavelengths ($\lambda> 170$~\AA).  Nevertheless, if the comparison had been made in $f_{\lambda}/f_{\rm bol}$, rather than $f_{\lambda}$, B would be systematically $\sim$3 times brighter at all the wavelengths.
}
\end{figure}

\clearpage
\begin{figure}
\figurenum{5}
\figcaption[]{Differential emission measure models of low and high states of AB (middle panels),
and comparisons of simulated and observed line strengths (flanking panels: left for high-FIP C, O, N, and Ne; right for low-FIP Mg, Si, Ca, Fe, and Ni).  Dot-dashed curves in DEM frames taken from neighboring panel below, for comparison.  Observed fluxes were normalized to the bolometric luminosity, $f_{\rm bol}$, of each star; and to $P_{\rm tot}$, the radiative power summed over $\log{T}$.  First normalization removes the bias of different stellar sizes; second allows features of very different intrinsic emissivity to be compared on a common scale.  Normalized observed fluxes (larger symbols) are plotted at an emissivity and DEM weighted formation temperature.  Calculated fluxes (smaller black symbols) were similarly normalized.  Down arrows indicate 3\,$\sigma$ upper limits.  Numbers in upper right hand corners of low-FIP panels indicate an abundance enhancement applied uniformly to those species.  Note that the DEM models are similar below about 1~MK, but deviate strongly at higher temperatures, with high-state variants showing more hot material than the low states.
}
\end{figure}

\clearpage
\begin{figure}
\figurenum{5}
\vskip 0mm
\hskip -12mm
\includegraphics[scale=0.75,angle=90]{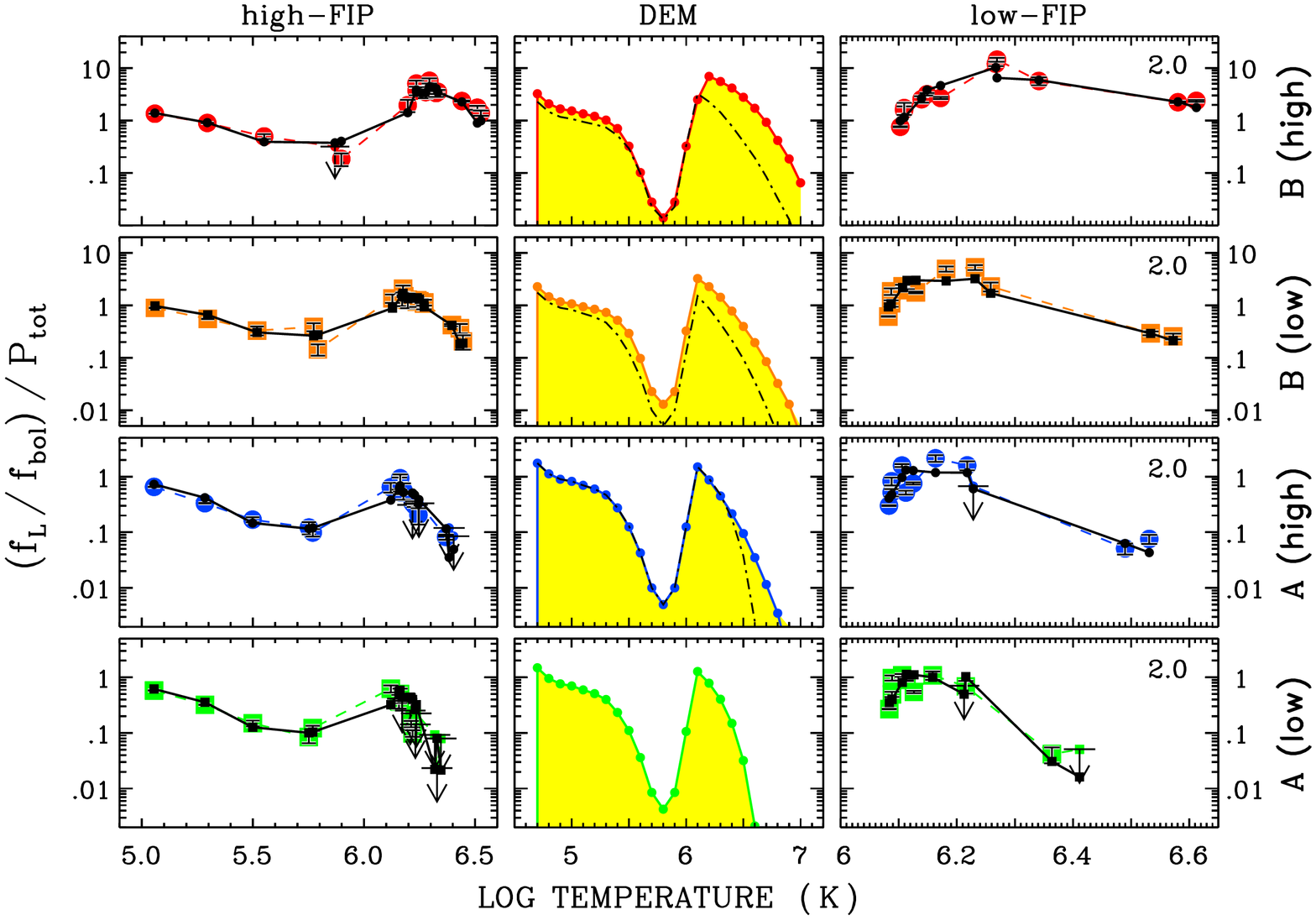} 
\figcaption[]{}
\end{figure}

\clearpage
\begin{figure}
\figurenum{6}
\vskip 0mm
\hskip -2mm
\includegraphics[scale=0.825,angle=90]{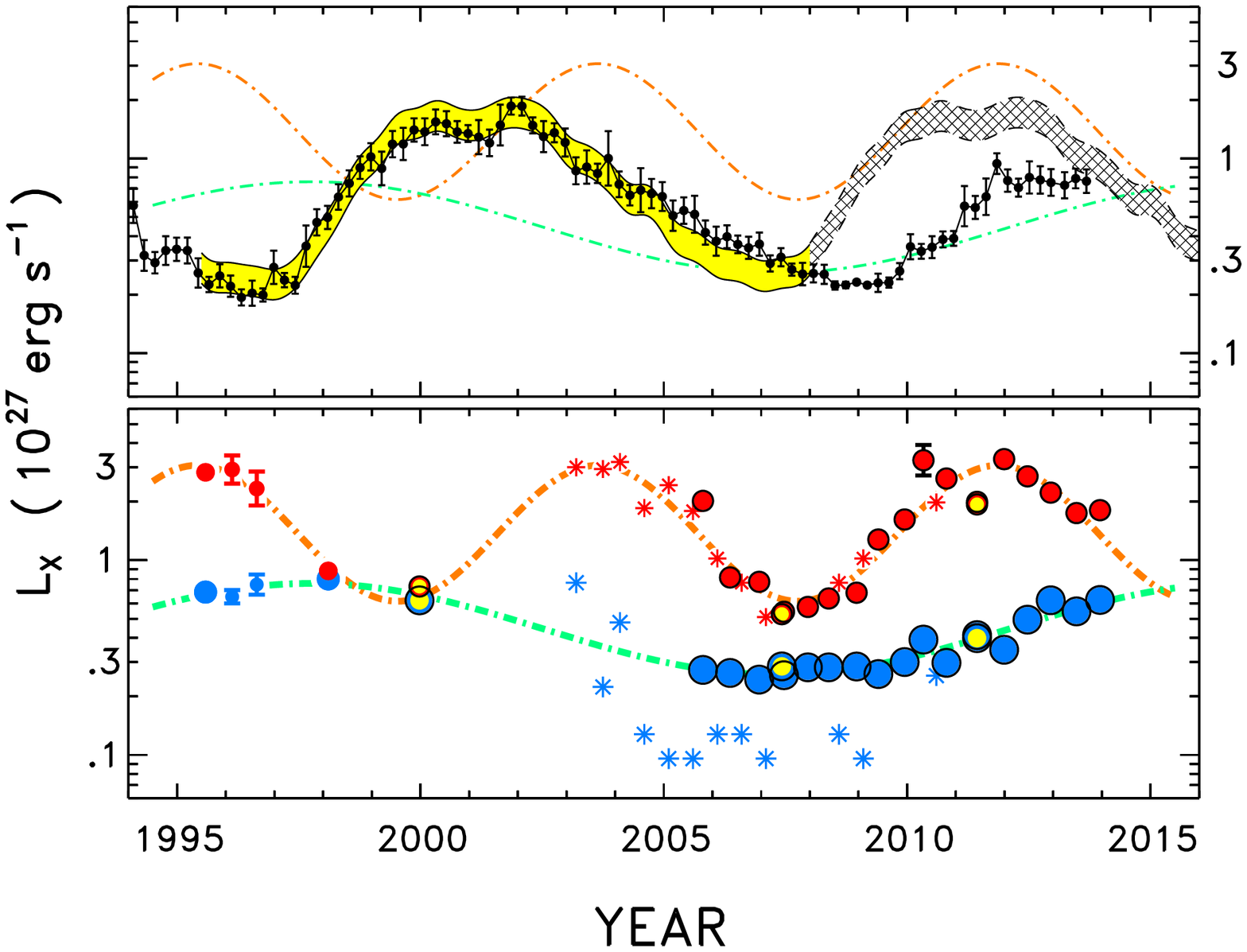} 
\vskip 0mm
\figcaption[]{{\em Upper Panel}-- Solar 0.2--2~keV luminosities, 81~day averages (three rotations), over Cycles 23 and 24 (black dots); error bars indicate 1\,$\sigma$ standard deviations of the daily measurements in each bin.  Three-cycle average is shaded yellow, projected into future as dashed curves and dark hatching (see Ayres [2009] for details).  Cycle 23 showed an unusual extended minimum.
{\em Lower Panel}--  Post-2000 solid dots depict HRC fluxes of $\alpha$~Cen: blue for solar-type primary; red for K-type secondary.  Pre-2000 dots represent four {\em ROSAT}\/ HRI epochs.  Yellow dots mark LETGS exposures.  Asterisks are reported {\em XMM-Newton}\/ X-ray luminosities of AB, scaled by $\sim$1.28 to match the apparent {\em Chandra}\/ cycle of B.  Dot-dashed curves, repeated in all panels, are schematic log-sinusoidal fits to {\em Chandra}\/ and {\em ROSAT}\/ for A, and including also scaled {\em XMM-Newton}\/ for B.  
}
\end{figure}

\end{document}